%% file: main.tex
\title{Thesis title}
\documentclass[
12pt, 
oneside, 
english, 
doublespacing,
headsepline, 
]{MastersDoctoralThesis} 

\usepackage{booktabs} 
\usepackage{amsmath}
\usepackage{algorithm}
\usepackage{arevmath}
\usepackage[noend]{algpseudocode}
\usepackage{palatino} 
\usepackage[]{datetime}
\usepackage[sortcites,backend=biber,style=numeric-comp,sorting=nyt]{biblatex}
\usepackage{adjustbox,lipsum}
\usepackage{makecell}
\usepackage{colortbl}
\usepackage[T1]{fontenc}
\usepackage{float}
\usepackage{array}
\usepackage{pdfpages}
\usepackage{attachfile2}
\usepackage[toc,page,title,titletoc]{appendix}
\usepackage{lipsum}
\usepackage{longtable}

\usepackage{tabularx}

\usepackage[T1]{fontenc}
\usepackage[utf8]{inputenc}
\usepackage{lmodern}
\usepackage[autostyle]{csquotes}
\usepackage{hyperref}
\usepackage[toc,page]{appendix}
\usepackage{multirow}
\usepackage{graphicx}

\usepackage{xspace}

\def \new #1{\color{black}#1\color{black}}
\newcommand{\remove}[1]{}
\newcommand{\etal}{\textit{et al}. }

\DeclareMathOperator*{\argmax}{arg\,max}
\DeclareMathOperator*{\argmin}{arg\,min}

\definecolor{darkpastelgreen}{rgb}{0.01, 0.75, 0.24}
\definecolor{ufogreen}{rgb}{0.24, 0.82, 0.44}
\definecolor{shamrockgreen}{rgb}{0.0, 0.62, 0.38}

\newtheorem{example}{Example}
\newtheorem{definition}{Definition}

\title{Deployment Optimization of IoT Devices through Attack Graph Analysis} 

\supervisor{\textsc{Dr. Rami Puzis} $\&$\\ \textsc{Dr. Asaf Shabtai}}
\examiner{} 
\degree{Master of Science} 
\author{\textsc{Noga Agmon}} 
\addresses{} 

\university{Ben-Gurion University of the Negev} 
\department{Department of Software and Information Systems Engineering} 
\faculty{Faculty of Engineering Sciences} 

\newdateformat{monthyeardate}{\monthname[\THEMONTH], \THEYEAR}
\bibliography{bibliography} 
\makeatletter
\def\@makeappendixhead#1{%
  \null\vfill%
  {\parindent \z@ \centering \normalfont
    \ifnum \c@secnumdepth >\m@ne
      \if@mainmatter
        \huge\bfseries \@chapapp\space \thechapter
        \par\nobreak
        \vskip 20\p@
      \fi
    \fi
    \interlinepenalty\@M
    \Huge \bfseries #1\par\nobreak
    \vfill
    \clearpage
  }}
\g@addto@macro\appendices{\let\@makechapterhead\@makeappendixhead}
\makeatother

\DeclareDocumentEnvironment{abstract}{ O{} }{
  \checktoopen
  \tttypeout{\abstractname}
    #1
  \thispagestyle{plain}
  \begin{center}
    {\huge\textit{\abstractname} \par}
  \end{center}
}

\begin{document}


\hypersetup{
    citecolor=black,
    linkcolor=black,
    filecolor=black,      
    urlcolor=black,
}


\pagestyle{plain} 

\begin{titlepage}

\begin{center}

\vspace{0.2cm}
{\scshape\large\univname\par}\vspace{0.1cm} 
{\scshape\large\facname\par}\vspace{0.1cm}
{\scshape\large\deptname\par}\vspace{1.0cm}
\includegraphics{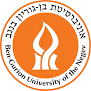}\\[0.5cm] 

\HRule \\[0.4cm] 
\vspace{0.4cm} {\Large \bf Deployment Optimization of IoT Devices through Attack Graph Analysis}
\HRule \\[0.5cm] 
 
\begin{minipage}[t]{0.4\textwidth}
\begin{flushleft} \large
\emph{Author:}\\
{\authorname} 
\end{flushleft}
\end{minipage}
\begin{minipage}[t]{0.4\textwidth}
\begin{flushright} \large
\emph{Supervisors:} \\
{\supname} 
\end{flushright}
\end{minipage}\\[2cm]
 

 
Thesis submitted in partial fulfillment of the requirements \\
for the M.Sc. degree \\[0.5cm] 
\large \monthyeardate\today 

\vfill
\end{center}
\end{titlepage}



\begin{center}

\vspace{0.2cm}
{\scshape\large\univname\par}\vspace{0.1cm} 
{\scshape\large\facname\par}\vspace{0.1cm}
{\scshape\large\deptname\par}\vspace{1.0cm}

\HRule \\[0.4cm] 
{\huge \bf \par}\vspace{0.4cm} Deployment Optimization of IoT Devices through Attack Graph Analysis
\HRule \\[0.5cm] 

\textsc{\large Supervisors: \supname}
\\[4.0cm]
\end{center}

\begin{minipage}{11cm}
\begin{flushleft}
Author: Noga Agmon\\
Supervisor: Dr. Rami Puzis\\
Supervisor: Dr. Asaf Shabtai\\
Chairman of Graduate Studies Committee: Dr. Rami Puzis\\
\end{flushleft}
\end{minipage}
\begin{minipage}{5cm}
\begin{flushright}
Date: September 11, 2019\\
Date: September 11, 2019\\
Date: September 11, 2019\\
Date: September 11, 2019\\
\end{flushright}
\end{minipage}
\\[4cm]
\begin{center}
\large \monthyeardate\today 
\end{center}

 







\begin{abstract}
\addchaptertocentry{\abstractname} 

The Internet of things (IoT) has become an integral part of our life at both work and home.
However, these IoT devices are prone to vulnerability exploits due to their low cost, low resources, the diversity of vendors, and proprietary firmware. 
Moreover, short range communication protocols (e.g., Bluetooth or ZigBee) open additional opportunities for the lateral movement of an attacker within an organization. 
Thus, the type and location of IoT devices may significantly change the level of network security of the organizational network.
In this work, we quantify the level of network security based on an augmented attack graph analysis that accounts for the physical location of IoT devices and their communication capabilities.
We use the depth-first branch and bound (DFBnB) heuristic search algorithm to solve two optimization problems: Full Deployment with Minimal Risk (FDMR) and Maximal Utility without Risk Deterioration (MURD).
An admissible heuristic is proposed to accelerate the search. 
The proposed method is evaluated using a real network with simulated deployment of IoT devices. The results demonstrate (1) the contribution of the augmented attack graphs to quantifying the impact of IoT devices deployed within the organization on security, and (2) the effectiveness of the optimized IoT deployment.


\end{abstract}

\keywordnames


\begin{acknowledgements}
\addchaptertocentry{\acknowledgementname} 
First and foremost, I thank my research supervisors, Dr. Rami Puzis and Dr. Asaf Shabtai, for their continuous guidance and support throughout this research.

Thanks to my boyfriend, Erez for his support in this journey. And my family
for providing certainty and comfort in times I needed it.
\end{acknowledgements}



{
  \hypersetup{linkcolor=black}
  \tableofcontents 
  \listoffigures 
  \listoftables 
}

\pagestyle{thesis} 


\input{macros}


\include{Chapters/all_chapters}





\printbibliography[heading=bibintoc]

\end{document}

%% file: macros.tex
\newcommand{\algo}{SMOBI\xspace}
\newcommand{\approach}{NO-DOUBT\xspace}
\newcommand{\fullapproach}{NOvel DOcUments Based aTtribution\xspace}

\newcommand{\fullalgo}{Smoothed Binary vector\xspace}

\newcommand{\bs}[1]{\textcolor{brown}{$\ll$\textsf{#1 --BS}$\gg$}}

%% file: Chapters/all_chapters.tex

\chapter{Introduction}
It is estimated that by 2020 more than 20 billion IoT devices will be deployed in the world~\cite{Meulen2017Gartner2016}. 
Most IoT products are not equipped to deal with security and privacy risks, which can turn them into the weakest link of organizational networks. 
The risk of IoT devices to the security of an organization is underestimated in many cases when an organization's IT department relies solely on network separation to isolate IoT devices from other IT assets. 
Such an approach disregards some of the unique properties of IoT devices, such as light or sound emissions, various sensors, and diverse communication protocols such as NFC, Bluetooth, ZigBee and LoRA, in addition to standard Wi-Fi. 
The advanced capabilities of IoT devices can be exploited by an attacker for lateral movement within an organization, shoulder surfing, and more, making them a valuable asset for an attacker.

With respect to hardening IoT security, most prior research focuses on the security of individual IoT devices~\cite{roman2011securing, liu2012research, zhang2014iot}, the security of an IoT protocol~\cite{Vaccari2017RemotelyNetworks, Zillner2015ZigBeeUgly, morgner2016all, wright2009killerbee, ronen2017iot}, or the the security of a network that consists solely of IoT devices~\cite{huang2014novel, skarmeta2014decentralized, zanella2014internet, Ge2016} (see Section~\ref{sec:rw_iot_dep} for more details). To the best our knowledge, there is no previous related research aimed at identifying the optimal (security risk-wise) deployment of devices within the physical space.
The location of an IoT device within an organization can have unintended effects on the network topology such as bridging between networks through short-range communication protocols (see Sections~\ref{sec:back-iot} and \ref{sec:back-src}). 
We use the following example to demonstrate the problem.

\begin{figure}[ht]
\centering
\includegraphics[width=0.75\textwidth]
{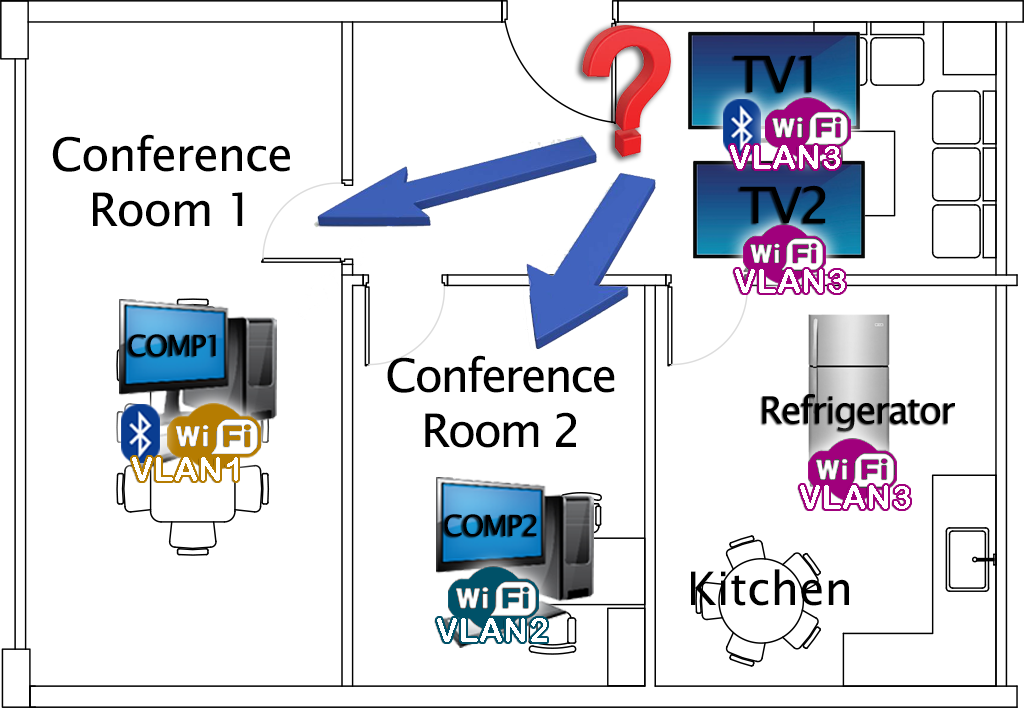}
\caption{Illustration of the office in example~\ref{run_example}.Different colors of Wi-Fi represent different VLANs.}
\label{fig:example_setup}
\end{figure}

\begin{example}[h]
\label{run_example}
Assume, for example, an office with two conference rooms and a kitchen (Figure~\ref{fig:example_setup}).
Each conference room has a computer ($COMP1$ and $COMP2$) connected through Wi-Fi to two different VLANs ($VLAN1$ and $VLAN2$ respectively). $COMP1$ also has Bluetooth. A smart refrigerator in the kitchen is connected to $VLAN3$ and has Internet connectivity. All other IoT devices in the office are connected to $VLAN3$ as well. The office purchased two televisions ($TV1$ and $TV2$) to replace the old projectors in the conference rooms. Both televisions are connected to $VLAN3$ via Wi-Fi; 
TV1 is also equipped with Bluetooth.

\textbf{Should we install $TV1$ in Conference Room 1 and $TV2$ in Conference Room 2 or vice versa?} 
To answer this question assume, for example, that unbeknownst to the organization, a sophisticated malware has managed to infect one of the computers in the organizational network.
Further, assume that the malware is equipped with the necessary exploits to hop between devices in the office. If $TV1$ is placed in Conference Room 1, the attacker could take advantage of the fact that both $TV1$ and $COMP1$ have Bluetooth and create an attack path to the refrigerator. However, if $TV1$ is placed in Conference
Room 2 this attack path will no longer be available to the attacker.
\end{example}

The risk of potential multi-step attacks such the one described in Example~\ref{run_example}  can be estimated using attack graphs~\cite{phillips1998graph,ou2006scalable}.
An attack graph is a model of a computer network that encompasses computer connectivity, vulnerabilities, assets, and exploits. It is used to represent a collection of complex multi-step attack paths (hereafter referred to as \emph{attack plans}) and can be used to assess and quantify security risk (see Section~\ref{sec:back-ag} for more details). 

In this research, the proposed method augments attack graph analysis to account for the physical location of IoT devices and their communication capabilities. (see Section~\ref{sec:iot_ag}).
Relying on the new attack graphs, we quantify the risk of adding an IoT device to a given network and show that 
it may increase by 22\% due to the deployment of only six IoT devices in a small to medium sized enterprise.

We also optimize the deployment of IoT devices in order to reduce the negative security implications of such deployment (see Section~\ref{sec:deployment_opt}). 
Two optimization problems are presented: 
the Full Deployment with Minimal Risk (FDMR) problem where all required IoT devices should be deployed with minimal security implications and the Maximal Utility without Risk Deterioration (MURD) problem where the maximal number of IoT devices should be deployed without increasing the security risk of the network. 
We use depth-first branch and bound (DFBnB) heuristic search algorithm to solve both optimization problems and suggest an admissible heuristic function to accelerate the search.
Our experiments show that optimal deployment of IoT devices can reduce the number of possible attack plans by 34\% (see Section~\ref{sec:evaluation}).

\chapter{Background} 
\label{sec:background} 
\section{Attack Graphs}
\label{sec:back-ag}

An attack graph is a model of a computer network that encompasses computer connectivity, vulnerabilities, assets, and exploits~\cite{phillips1998graph,ou2006scalable}. 
Attack graphs are used to represent collections of complex multi-step attack scenarios traversing an organization from an initial entry point to the most critical assets. 
By analyzing the attack graph, a security analyst can assess the risks of potential intrusions and devise effective protective strategies. 
The attack graph analysis methodology contains three main stages: (1) network and vulnerability scanning, (2) attack graph modeling, and (3) attack graph analysis. 

In the first stage, the Nessus vulnerability scanner~\cite{beale2004nessus} is used in order to map the vulnerabilities of all of the hosts in the organization. 
Connectivity between the hosts can be identified manually by system administrators based on the organizational network topology and firewall configurations. 
Nessus, Nmap, or other network scanners can aid in the connectivity assessment process. 

Network connectivity and vulnerability reports are processed by MulVAL~\cite{ou2005mulval} to generate an attack graph representation in planning domain definition language (PDDL). 
An attack graph consists of privilege nodes, exploit/action nodes, and fact nodes. In an attack graph, a privilege node represents the information gained or the access privileges that the attacker obtains (represented by triangles in the graph). An exploit/action node represents the action the attacker needs to exploit a vulnerability (represented by ovals). The edges of exploit nodes are for preconditions and postconditions of the exploit. A fact node represents a network condition that needs to exist in order for the attacker to exploit the vulnerability (represented by rectangles).
To gain a privilege, an attacker needs to execute one of the actions leading to it (logical OR). To use an exploit, the attacker needs all of the privileges and the facts that lead to the exploit (Logical AND). An exploit node needs all of these preconditions leading to it to be executed, and once executed, the attacker gains all of the postconditions the exploit node leads to~\cite{ou2006scalable,Sawilla2008IdentifyingGraphs,Ammann2002ScalableAnalysis,noel2014metrics}.

\begin{figure}
\centering
\includegraphics[width=0.75\textwidth]
{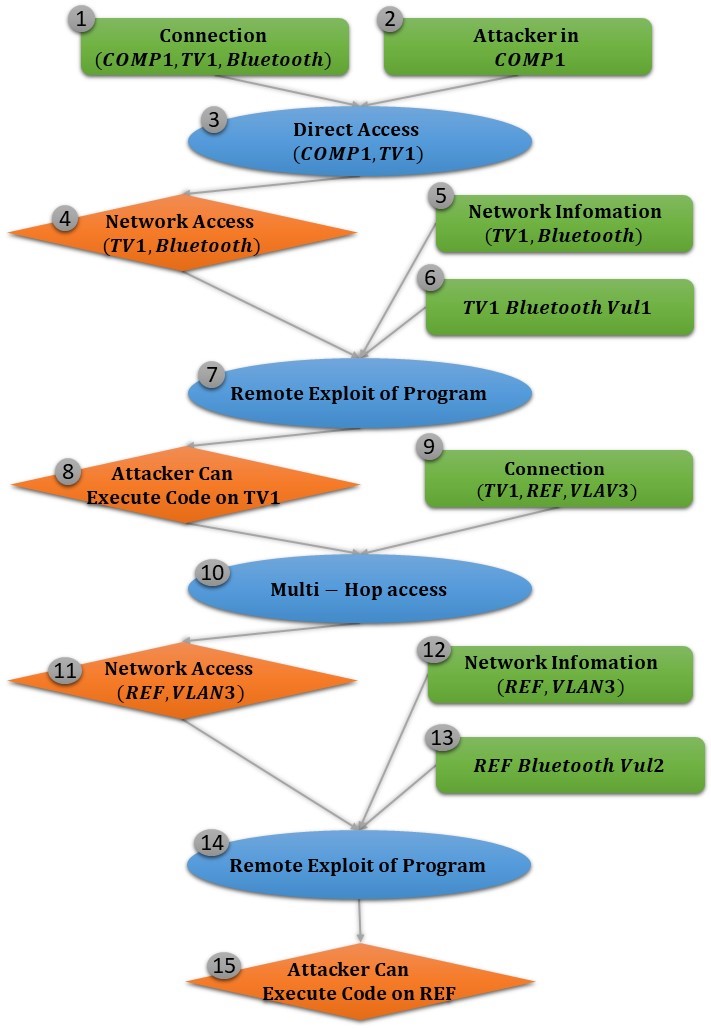}
\caption{Attack Graph of Example~\ref{run_example_ag}. Exploit/action nodes are represented by blue ovals; fact nodes are represented by green rectangles; privilege nodes are represented by orange diamonds.} 
\label{fig:attack_graph}
\end{figure}

\begin{example}
\label{run_example_ag}
Figure~\ref{fig:attack_graph} presents an abstract attack graph of the situation described in example~\ref{run_example}.
At the top of the figure, two fact nodes (nodes 1 and 2) that represent two facts of the system can be seen (green rectangles).
Access between $COMP1$ and $TV1$ can only created if these two conditions exist, as can be seen from the blue oval, which represents an exploit node (node 3). 
This access allows the attacker to use the Bluetooth connectivity of $TV1$, as represented by the orange diamond (node 4), meaning that the attacker can obtain control of $TV1$ via $COMP1$.
\end{example}

Following the construction of an attack graph, the graph's PDDL representation can be used as a domain model for variety of planners. 
A typical task is finding the optimal attack plan or estimating the likelihood of a successful attack given the attack graph of an organization~\cite{Singhal2011SecurityGraphs, Wang2008AnMetric, noel2014metrics}.
Consequently, attack graphs can be used for hardening network security through a variety of attack graph optimizations~\cite{islam2008heuristic, abadi2006ant, polad2017attack, noel2003efficient}.

\section{Internet of Things}
\subsection{Security in the Internet of Things}
\label{sec:back-iot}
Traditional security solutions such as firewalls, IDSs, anti-viruses, and software patches are not suitable for IoT devices.
The four major reasons for this are~\cite{Yu2015HandlingDevices}: 
(1)~Types of policies: a single app may use several IoT devices, communicating explicitly (e.g., via Wi-Fi or Bluetooth) or implicitly (e.g., an IoT light bulb can be triggered by an IoT light sensor). 
The outcome is a complex and dynamic network which can be hard to secure using a single security policy (e.g., with firewalls).
(2)~Signatures and anomalous behavior recognition: some security methods store anomalies and signatures on the device to recognize and detect threats. 
Due to the diversity of IoT devices and manufacturers, these methods will be inadequate, mainly because of the constant need to update and maintain the device to support these tools.
(3)~Enforcement mechanism: IoT devices have low computation abilities, low power consumption, and do not run full-fledged operating systems. Most common security methods need all of the above to operate and therefore are impractical to implement on IoT devices.
(4)~Unsupported devices: the longevity of IoT devices can lead to deployed devices that vendors no longer support. In that way, vulnerable devices (with default passwords or unpatched bugs) can remain in the organization.

Moreover, the competitive IoT device market compels vendors to try and get their products out as fast as they can, prioritizing functionality and the user experience, and ignoring the security aspect. 
In general, most products hardly deal with security and privacy risks, making them the weakest link in terms of security and the target of attackers interested in breaking into networks and harming systems or leaking information~\cite{Yu2015HandlingDevices}. 
Thus, despite the fact that security was recognized as a central issue of the IoT market as early as 2011 by Bandyopadhyay \etal~\cite{Bandyopadhyay2011InternetStandardization}, it still continues to remain a challenge today. 

\subsection{Vulnerabilities}
Security experts need to gather information about the vulnerabilities of devices in their organizations ~\cite{Arnaert2016ModelingSecurity}. To do so, they can use a specialized search engine, like Shodan or Censys, to try and find the answer. 

Shodan is a search engine created by John Matherly in 2009. It examines devices ports, take the resulting banners and indexing the corresponding IP address. This is possible as a result of many IoT devices (routers, printers, cameras etc.) broadcasting their presence over the network, some even include their device type and firmware version. This information is crossed with known exploits databases such as NIST and Mitre. Since in the Shodan database almost 4 million IP addresses are mapped, one of the problems of these search engines is that it can return thousands of results due to the amount and different kinds of devices and the quantity of the known vulnerabilities. The paper proposes an ontology that helps to decrease the number of results, find the most relevant results more easily and therefore may be useable to non-security experts by reducing the complexity of the results.
But those search engines can also be used for attacks.

\cite{Markowsky2015ScanningThings} show how they searched for “router” in the Shodan Exploits database and found an exploit that refers to Cayman router. After that they searched for the specific kind of router in Shodan database and obtained the IP address of many Cayman routers. They then used the exploit in the routers for their purpose, conducting a denial-of-service (DDoS) attack for example.

For a year~\cite{Cui2010ADevices} scanned regularly parts of the Internet to find trivially vulnerable embedded devices. They found that 540,000 devices, which are 13\% of all the discovered devices that came up in the scanning, were identified as configured with a root password. The devices were from 144 countries, almost 18,000 unique enterprises and of a variety of kinds (routers, printers etc.). From those devices 96\% remained accessible even after 4 months. In 2012, after scanning almost the entire network, the hacker created "Carna botnet" discovered 1.2 million devices that enabled login with default (e.g. root:root, admin:admin) or empty credentials~\cite{Anonymous2012Internet2012}. By infecting devices so they will also scan the internet, a distributed scan over all the IPv4 took only an hour.

In their research~\cite{Patton2014UninvitedIoT} also tried to measure the amount of vulnerable devices in the world. Their results show that the vulnerability rates range from 0.44\% to 40\%. On a global scale, 0.44\% means 4,400 vulnerable devices for every million IoT devices deployed (as mention above, for the present there are 8.4 billion devices deployed).
Any device that is connected to the internet and does not have a default and unique password is probably installed with a default login credentials and as a result of that, at risk of being vulnerable to attacks.

NAT networks are an exemption to this  since devices that use NAT can be secure, unless hackers compromise a system with access to the NAT, and then those devices are no longer safe from attacks.

\subsection{Attacks}
It’s not that there were more attacks on IoT than on PCs, it’s the relative effortlessness needed to attack IoT that is the problem.
Since the finding of the vulnerable devices as mentioned above, things didn’t improve. 

In 2014, 750,00 spam emails were sent by a smart fridges botnet~\cite{BBC2014FridgeGadgets}. In that same year, a DDos attack conducted using thousands of IoT devices such as home routers, paralyzed Microsoft and Sony’s gaming networks~\cite{Krebs2015LizardRouters}. In 2016, one of the most famous attack occurred, "Mirai botnet"~\cite{Stavrou2017DDoSIoT}. The Mirai mainly relies on hacking IoT devices by guessing the login credentials, using brute force of small, common username-password pairs. The attack was two DDoS that happened almost simultaneously, the first one against the website of security consultant Brian Krebs (620 Gbps of traffic) and the second against French webhost and cloud service provider (1.1 Tbps of traffic). After the first attack, the creators released the source code of Mirai. This triggered more attacks, such as the one on October 2016 against service provider Dyn, which took down hundreds of websites, among them Twitter, Netflix, GitHub and Reddit. These are only a few of the attacks performs, and the list can go on. 

As opposed to PCs, IoT are connected 24/7 to the network, have no antivirus or IDS as a defense and most of them have easy accessible credentials (e.g. weak password)~\cite{Pa2015IoTPOT:Compromises}. They may lack computational power, but compensate in numbers. The neglection of the security aspect by users and the manufactures alike, increases the chances of finding large numbers of vulnerable IoT, so from the attacker point of view, attacks on IoT devices seems really worthwhile.

\subsubsection{Solutions}
\cite{Pa2015IoTPOT:Compromises} analyzed a darknet for telnet based scans, founding that a majority of IoT have already been involved in attacks. These results motivated the authors to propose a IoTPOT, a novel honeypot that interacts in Telnet protocol with a variety of IoT devices. This honeypot mimics many IoT with different CPU architectures and has the ability to learn new, unknown commands. To analyze the malwares from the IoTPOT, the authors also present IoTBOX, a sandbox that can run malwares from different CPU architectures. They found four malware families, of which DoS attacks were the goal of three of them. The results from the experiments were good, and since then more papers tried to improve this work.

After showing how to hack smart lights bulbs~\cite{Ronen2016ExtendedLights} suggest some security measures that must be considered in the design, implementation and integration stages of IoT devices. First, the authors argue for the need of implementing standard security protocols, such as TLS. Among other benefits, this protocol enables the generation of random passwords as credentials for devices. Second, the authors argue for the need to try to limit the Application Programming Interface (API) to a minimum . If the user doesn’t need some of the functionality, it’s better not to have this functionality option at all, as it can be used for undesirable things. In the case of the hacked smart light bulbs for example, there was no need for so many  brightness levels. Finally, it’s important to consider how to integrate the IoT devices. In the smart light example, because those bulbs can be implemented into city's power grids, it’s better to do so only after separating them from the network in order to avoid attacks, such as blackouts or the one the authors performed.

Today, most of the research addressing the security problems of IoT, is about the scale and consequences of these problems without suggesting real solutions.~\cite{Yu2015HandlingDevices,MahmoudInternetMeasures,Gazis2015ShortArchitectures} each suggest a design or an architecture of IoT network that addresses some of these issues, sometimes in a general way and sometimes in a more practical way. To this date, to my knowledge, there is no acceptable and practical solution that can solve even some of the security issued mentioned above.

\section{Short Range Communication Protocols}
\label{sec:back-src}
When connecting a device to a network it is possible to use two categories of networking technologies. 
The first and simplest category is to connect using standard existing network technologies such as Wi-Fi and Ethernet. The second category is to connect using different wireless technologies that are more suitable for some devices, e.g., technologies that are more appropriate for devices that require low energy consumption protocols.
These protocols are short-range communication protocols, due to their requirement for short proximity in order to perform a connection.

Currently, in the second category there are several communication methods that can be used, including: ZigBee, Z-Wave, Powerline, Bluetooth 4.0, and other radio frequency protocols, but no standard protocol exists. Both Z-Wave and ZigBee are considered secure, but implementation flaws and manufacturer mistakes make them vulnerable~\cite{InsecuritySymantec}. 

In our research, we focus on ZigBee and Bluetooth, since they are ones  of the most common wireless technologies used to connect IoT devices. 
First, we start with the ZigBee protocol, which guarantees low power consumption and a two-way, reliable, wireless communications standard for short-range applications. 
It is open-source and has advantages such as easy deployment and global usage.

The ZigBee protocol was created with security considerations in mind, but consumer demand for cheap devices with long life expectancy often caused vendors to sacrifice security, which led to poor implementation of the protocol~\cite{Zillner2015ZigBeeUgly}; this, in turn, led to major security issues such as data compromising or information sniffing~\cite{Vaccari2017RemotelyNetworks}.
For example, Vaccari \etal~\cite{Vaccari2017RemotelyNetworks} focused on the security aspects of the ZigBee protocol. 
The study identified important security issues and presented an attack on the protocol which enabled the attacker to compromise the data transferring in the network.  
Morgner \etal~\cite{morgner2016all} described a novel attack that shows that the ZigBee Light Link standard is insecure by design. 
Wright \etal~\cite{wright2009killerbee} published KillerBee, a penetration testing tool which allows ZigBee traffic to be sniffed and analyzed.
Ronen \etal~\cite{ronen2017iot} found a major bug in the ZigBee protocol in Philip Hue smart lamps. 
They were able to perform an over-the-air firmware update, thereby infecting the lamp with a worm that can spread to any of the lamp's neighbors.

Bluetooth was developed by a group called the Bluetooth Special Interest Group (SIG) in May 1998. Today, a lot of smartphones, sports devices, sensors, and medical devices have Bluetooth. 
The protocol become widely used because of its low cost and low power consumption.

In~\cite{ryan2013bluetooth}, techniques were presented for eavesdropping on devices using Bluetooth. 
An extended review of Bluetooth threats and possible attacks was performed by Minar~\etal, Sandya~\etal and Dunnin ~\cite{minar2012bluetooth, sandhya2012analysis, dunning2010taming}, and recently, Cope \etal~\cite{cope2017investigation} investigated the currently available tools to exploit vulnerabilities in Bluetooth. 
In conclusion, many Bluetooth versions that are in use today, have a wide variety of security vulnerabilities.

In addition to the security issues, the number of communication protocols in an IoT device can also influence the security of the device. 
If such a device is compromised by an attacker that has hacked into one of its communication protocols, the hacker can take advantage of the compromised device and use the other protocols as entry points to the network~\cite{Ge2017SecurityDevices}.

Of all the above, short-range communication protocols are another aspect of IoT devices that make them insecure compare to regular~hosts.

\section{Heuristic Search }
\label{sec:back-search}
Heuristic search is a family of techniques used to solve difficult problems in artificial intelligence (AI). 
In this case, each problem is represented by states, where each state represents the current condition of the problem. 
Each problem also has a starting state and one or more goal states. 
A search space is the environment in which a search takes place, where the purpose of the search is to find a path from the start state to one of the goal states in the search space. 
Each solution represents by one goal state. 
The quality of the solution is measured by the cost of the goal state.
Search algorithms make a distinction between minimum and maximum problems. 
In a minimum problem, we want to find the solution with the lowest cost, and in a maximum problem the highest cost solution is desired. 
Most problems are minimum problems, e.g., we want the cheapest or the fastest solution. If not stated differently, we are referring to a minimum problem.
In our research, we use the depth-first branch and bound algorithm~\cite{Korf2010ArtificialAlgorithms,Zhou2006Breadth} which uses a heuristic function to solve the problems more efficiently.

\subsection{Heuristic}
A heuristic is an estimation of the cost of the path from node $n$ to a goal node. The heuristic function is used to steer the search algorithm in the direction of the goal.
In an informed way, heuristics help the algorithm guess which child out of all of the node's children will lead to the goal.


\subsection{Admissible Heuristic} 
If, for any $n$, a heuristic function never overestimates the cost of the best path from node $n$ to a goal node, then the function is referred to as an \emph{admissible heuristic function}. 
Note that in a maximum problem (where we want the solution with the maximum cost) it is the opposite, i.e., a heuristic function that never underestimates the cost of the best path. 

For example, suppose we try to find the shortest path from point A to point B. 
This is a minimum problem where the cost is the distance between the two points. 
A good admissible heuristic function could be the aerial distance, because the actual path between both points will never be lower than the aerial distance.

In most search algorithms, \new{as well as in DFBnB algorithm, } one of the most important conditions for a heuristic function is that it should be admissible.

\subsection{Algorithms}
We will review three common searching algorithms, $A*$, $IDA*$ and $DFBNB$~\cite{Korf2010ArtificialAlgorithms}.

\textbf{A* Algorithm. } The basic algorithm is A*. A* is best-first search algorithm that find the optimal path if the heuristic function is admissible. His advantage is that he expands the fewest number of nodes. His main drawback is the memory requirement, like any best-first search algorithms, and therefore cannot really solve most of the complex problems.

\textbf{IDA* Algorithm. } IDA* is Iterative-deepening A* and he eliminates the memory problem of A*, without harming the optimality. Each iteration of the algorithm is a depth-search that keeps track of the cost that is calculated as in the A* algorithm. When a cost is higher than a certain threshold, it is cut off and the search backtracked and then continued. The initial threshold is the heuristic estimate of the start node, when in each successive iteration its increased by the lowest cost that was pruned in the previous iteration. IDA* memory requirements are linear in the maximum depth search, and his solution is optimal if the heuristic function is admissible. He is asymptotically optimal in time and space over A* and easier to implement. IDA* is good for the sliding puzzles or Rubik's cube problems. 

\textbf{DFBnB Algorithm. } 
Depth-first branch and bound (DFBnB) is a Depth-first search algorithm~\cite{Zhou2006Breadth, Korf2010ArtificialAlgorithms}.
The algorithm is used to navigate through the search space and find the optimal solution.
During the search process, DFBnB maintains the best solution found so far. 
In order to perform pruning more frequently and thus accelerate the search process, DFBnB uses a heuristic function.
\new{
The cost of a node is defined by $f(x) = g(x)+h(x)$, where f(x) is the sum of (1) the current cost of the node $g(x)$, and (2) the heuristic function for the goal $h(x)$, which is the estimation of the cost from node $x$ to the goal. 
Thus, $f(x)$ estimates the lowest total cost of any solution path going through node $x$.
If a branch with a higher cost is found, it can be pruned so that there is no need to keep expanding towards it.
}

The algorithm returns an optimal solution with linear memory space, assuming the heuristic function is admissible.
This solution depends on the kind of problem (i.e., minimum or maximum),
which is determined by the cost of the goal state.


DFBnB is suitable for problems when the maximum search depth is known in advance or the search tree is finite, like in the traveling salesman problem (TSP) or like in our problem, because each node have excatly two children.
In this research we chose the heuristic algorithm DFBnB, as it fits our problem.

\chapter{Related Work} 
\label{sec:related_work}
\section{IoT Device Deployment}
\label{sec:rw_iot_dep}
There are several works regarding the deployment of IoT devices, but most of them do not consider the security aspect. 
For example, Huang \etal~\cite{huang2014novel} proposed a deployment scheme used to a achieve green networked IoT, while Skarmeta \etal~\cite{skarmeta2014decentralized} focused on privacy issues and Zanella \etal~\cite{zanella2014internet} focused on the IoT in smart cities.

Some of the research that refers to security analyzes single IoT devices but does not look at IoT devices as a deployment problem.
Liu \etal~\cite{liu2012research} tried to solve the problem of assessing the risk of a single IoT device, by proposing a dynamical risk assessment method inspired by an artificial immune system. 
Zhang \etal~\cite{zhang2014iot} and Roman \etal~\cite{ roman2011securing} reviewed security issues in the IoT in terms of the security of each device. 

There are a few works that refer to deployment and network security, but they do not take the combination of hosts (such as computers and servers) with IoT devices into consideration.
Mohsin \etal~\cite{mohsin2017iotriskanalyzer} argued that the likelihood of exploiting IoT vulnerabilities depends on the system configuration. 
The authors explained that various configurations derive from different devices, technologies, and connectivity, all of which serves the same goal but have different risk levels.
Santoso \etal~\cite{santoso2015securing} presented an approach to secure smart home systems in which IoT devices are deployed, and Abie \etal~\cite{abie2012risk} introduced a risk-based adaptive security framework for the IoT in health-care systems.
\new{Shivraj \etal~\cite{shivraj2017graph} argue that most of the existing risk assessment frameworks are not suitable for IoT devices, and suggested a new generic risk assessment framework for IoT systems. The authors refer to systems like smart cities, smart grid, etc. and base their approach on graph theory. }

The research mentioned above reflects the many challenges of IoT security. 
In this respect, our work is unique in two ways. 
First, it combines the security concerns of the IoT with workstations and servers, while taking into account the possible use of one to hack the other.  
Second, our network model is a generic network that can be suitable for a variety of scenarios and is not specific for a particular domain.

\section{Attack Graph Optimization}
\label{sec:rw_ag-opt}

\subsection{Attack Graph Representation} 
Attack graphs have been used to estimate the security risk score of organizational networks~\cite{Singhal2011SecurityGraphs, Wang2008AnMetric, noel2014metrics}.
These works try to overcome different problems and rely on multiple tools while presenting different models for risk analysis,
however the specific characteristics of IoT devices were not considered in these articles.
In all of this research, the structure of the regular IT network is analyzed, taking into account the vulnerabilities of workstations and servers. 
IoT devices introduce additional challenges to security risk modeling through attack graphs, such as the diverse physical locations, variety of short-range communication protocols, cyber-physical capabilities of the devices, mobility, etc. 

In this research, we augmented the attack graph model of an organization to consider locations and short-range communication of IoT devices, and we used the augmented attack graph model to optimize the deployment of IoT devices throughout the organization. 

\subsection{Risk Score} 
Wang \etal~\cite{Wang2008AnMetric} suggested an overall network security score by combining individuals' vulnerabilities regarding their relationship in attack graphs.
Singhal \etal~\cite{Singhal2011SecurityGraphs} defined the risk score as the likelihood of an attack which was derived from the likelihood of individual exploits. 
Noel \etal~\cite{noel2014metrics} described four families of metrics for measuring security risk in attacks graph. 
Every family was represented by one entry in a four-dimensional vector.   
The Euclidean norm of this vector was used as the overall risk score.
Gonda \etal~\cite{gonda2017ranking} computed the number of shortest plans in a planning graph derived from an attack graph as a way to measure the security of the network, and Swiler \etal~\cite{swiler2001computer} computed the set of near-optimal shortest paths to identify the most exploitable components in the network. Polad \etal~\cite{polad2017attack} used an attack graph to estimate the security of the network as the cost of the attack path that led to the goal.

\new{
Idika \etal~\cite{idika2010extending} used three risk assessments metrics, The Shortest Path metric, the Number of Paths metric, and the Mean of Path Lengths and combined them to determine between two networks which one is more secure. The authors claimed that each risk score by itself can lead to misleading results, and suggested an algorithm for combining the usage of attack graph-based security metrics. They used decision metrics when two metrics created a conflicts of which network is more secure. Their algorithm helps reaching a decision by creating a total ordering of priority on questions regarding these conflicts.
}

All the above risk scores can be used to optimize the IoT deployment once the attack graph definition has been augmented to take into account the IoT device specifications. 

\new{
In our research, we find, as did Idika\etal, that one metric cannot represent the risk of the network correctly, and suggested a simple way to combine different measures in order to compare the risk between different networks.
We adopt Gonda's approach and used planning graph to find all shortest paths in the graph. 
We combined their quantity, their length, the average number of exploits as well as the average number of privileges the attacker obtained to one risk score (see Section~\ref{sec:risk_score}).
}

\remove{
In this paper, we adopt Gonda's approach to measure network security and combined it with Polad's method, to include the length of the shortest plans, as well as their quantity (see Section~\ref{sec:risk_score}).
}

\subsection{Optimization Problems} 
Security risks can be reduced by patching vulnerabilities. 
However, it is not always possible to patch all vulnerabilities at once due to operational costs (patching often requires significant downtime). 
A variety of low cost network hardening approaches can be used to prioritize the vulnerabilities (e.g., \cite{noel2003efficient,jun2011minimum}). 
Islam \etal~\cite{islam2008heuristic} argued that most of these methods are not scalable.
They proposed heuristic algorithms to accelerate the patch optimization. Abadi \etal~\cite{abadi2006ant} used the ant colony optimization algorithm to detect a minimum critical set of exploits. 
Polad \etal~\cite{polad2017attack} examined the effect of adding fake vulnerabilities in an attack graph and used combinatorial optimization in order to find optimal assignment of these vulnerabilities.
Almohri \etal~\cite{almohri2016security} used sequential linear programming in attack graphs to find the optimal placement of security products (e.g., a host-based firewall) across a network. 
The authors used a probabilistic model which uses Bernoulli and transformed the attack graph into a system of linear and nonlinear equations.
Noel \etal~\cite{noel2008optimal} used attack graph to optimize the placement of intrusion detection system (IDS) sensors to allow monitoring malicious activity on critical paths.

In this work we present a different optimization problem of optimizing the set of IoT devices to be deployed throughout an organization with minimal implications to the network security.

\section{IoT in Attack Graphs}
\label{sec:rw_iot_ag}
Very little work has been performed on attack graphs that consist of IoT devices. 
The first research performed in this area was conducted by Ge \etal~\cite{Ge2016} who used attack graphs in conjunction with IoT devices.
However, the network used consisted only of IoT devices, most of which were the same kind of device. 
The network topology was fixed, small, and relatively uncomplicated. 
The authors proposed a framework for IoT device security modeling with the aim of presenting all possible attack paths in the network, evaluating the security level, and assessing the effectiveness of different defense strategies.

In a later work, Ge \etal~\cite{Ge2017SecurityDevices} noted that some IoT devices use more than one communication protocol. 
The writers argued that if such a device is compromised by hacking into one of the communication protocols, the hacker can take advantage of it and use the other protocols as entry points to the network. 
The paper used HARMs (hierarchical attack representation models), which are models of attack graphs used, to improve scalability~\cite{hong2012harms}.
The authors presented a real scenario and showed how an attacker can take advantage of it.
In the scenario, some devices have both Wi-Fi and ZigBee communication protocols. 
Also present are smart devices such as a tablet and TV that can connect to a Philips Hue lighting system (Hue Bridge) by Wi-Fi. 
This lighting system also has ZigBee which allows it to control smart light bulbs in the house.
By exploiting the tablet that runs the Hue application, an attacker can gain control of the Hue Bridge system and use it to control all of the smart lights. 
The authors noted that the lighting hub can consist of any other smart hub, and the scenario can also be used to hack into any smart device, not only light bulbs.

\new{George \etal\cite{george2018graph} used attack graphs to implement risk assessment framework for Industrial IoT systems (IIoT). Such systems can be found in health-care, agriculture, transportation, etc. The authors noted the many vulnerability in IoT devices and claimed that it may lead an attacker to intrude the system using a weak IoT device. Their framework was inspired by attack graphs, allowing them to proposed risk mitigation strategies to reduce the overall threat level in the network. }

Yi{\u{g}}it \etal~\cite{yiugit2019cost} proposed COBANOT, a heuristic-based cost and budget aware network hardening solution for IoT systems which uses compact attack graphs~\cite{chen2010scalable}.
This work is the first to use attack graphs in IoT systems for network hardening. 
However, their experiment included a small-scale attack graph that only consists of IoT devices. In addition, none of the unique characteristics of IoT devices, such as different protocols, mobility, physical proximity, etc. were considered.

Our research focuses on networks that combine all kinds of hosts and IoT devices. 
Also, our network's size is larger than the networks used in the research mentioned above.



\chapter{IoT Attack Graphs}
\label{sec:iot_ag}
\section{IoT Deployment}
In a typical organization, all hosts (workstations and servers) are connected to the organizational network via a wired or wireless connection. 
Let $H = \{ h_1, h_2,\ldots, h_z\}$ be the set of hosts that are part of the organization network.

In addition to the regular hosts, the organizational network may contain IoT devices.
Let $D = \{d_1, d_2,\ldots, d_m\}$ indicates the set of unique IoT devices.
Each IoT device $d_i$ has a unique identifier (usually an IP address).

IoT devices differ by their purpose and capabilities. 
For example, a refrigerator is capable of maintaining a low temperature while a smart TV is capable of showing high definition movies. 
We group IoT devices by type, e.g., refrigerator,  TV, camera, smoke detector, etc.
$T = \{t_1, t_2,\ldots, t_n\}$ is the set of all the IoT device types. 
We denote a set of all devices that are of type $t$ as $D(t)$ and a single device type $d$ as $t(d)$.
We assume that every IoT device is part of just one group.

Some IoT devices can only be deployed in specific predefined designated locations. 
For example, the kitchen is typically the designated location for a refrigerator, while large TV screens or projectors are found in meeting rooms. 
Some IoT devices such as cameras or smoke detectors may be deployed in many different locations throughout an organization.

\begin{definition}\textbf{Locations.}
\label{def:locations}

$L = \{l_1, l_2,\ldots, l_b\}$ indicates the set of unique location spots where IoT devices can be deployed. 
We denote the set of locations where an IoT device of a specific type $t\in T$ can be deployed as $L(t)\subseteq L$. 
In every location spot only one type of IoT devices can be deployed, meaning, $L(t)$ is defined such that the intersection of each pair of $L(t_i)$ sets are empty, $\cap_{t \in L(t)} = \emptyset$. Because a location spot must be associated with some type of IoT devices, the union of $L(t)$ is equal to $L$, $\cup_{t \in L(t)} = L$
\end{definition}

Organizations may have constraints about the deployment of IoT devices. 
We defined two main constraints for a device type $t$. 
The first one is the number of locations (out of the total locations available) that need to contain a deployed device of that type.
For instance, there are four possible locations for cameras in the hallway, but the organization only needs to deploy two of them.
The second constraint is the number of devices there are of each type. 
For instance, for one location in which a refrigerator can be deployed, there are three possible refrigerators that the organization can purchase.

\begin{definition}\textbf{Location Constraint.}
\label{def:locations_const}

Let $C$ be the set of all constraints. $C(t)$ is a three-tuple that represents a constraint for a type $t$, $C(t) = (L(t), n(t), D(t))$. 

$L(t)$ is the set of locations that an IoT device of a specific type $t$ $\in$ $T$ can be deployed (as defined in Definition~\ref{def:locations}).

$n(t)$ is the number of locations that needed to be deployed out of all locations in $L(t)$.

$D(t)$ is a set of all IoT devices that are of type $t$.
\end{definition}

An example of a constraint can be derived from example~\ref{run_example}. 
Suppose the organization has three possible locations in which a TV can be deployed ($L(TV)=\{l_{TV_1},l_{TV_2},l_{TV_3}\}$) but only needs to deploy a TV in two of these locations ($n(TV)=2$). 
In addition, there are four different televisions that can be deployed ($D(TV)=\{d_{tv1},d_{tv2},d_{tv3},d_{tv4}\}$). 
Formally, constraint $C(TV)$ would be defined as follow: 

$C(TV) = (\{l_{TV_1},l_{TV_2},l_{TV_3}\},2, \{d_{tv1},d_{tv2},d_{tv3},d_{tv4}\})$.

Assume that at most one IoT device can be deployed in each location $l\in L$.
The deployment of IoT devices is defined as a function $depl:D\rightarrow L\cup \{\bot\}$ which maps every device to a particular location. 
The special non-location symbol $\bot$ signifies that a device is not deployed.  
We say that a deployment is valid if it does not violate the constraints specified in Definition~\ref{def:locations_const}.

\begin{definition}\textbf{Valid Deployment.}

Let $depl:D\rightarrow L\cup \{\bot\}$ be a deployment of IoT devices. $depl$ is valid if $\forall_{d\in D}, depl(d)\in L(t(d))\cup\{\bot\}$.
\end{definition}

We denote $depl_{full}$ as a deployment that satisfies all constraints~$C$ and $depl_{empty}$ as an empty deployment with no IoT devices deployed.
Note that the condition $n(t) \le |D(t)|$ should be satisfied for full deployment to exist.

Many IoT devices deployed within an organization's premises will likely be able to communicate with nearby hosts via short-range communication (SRC) protocols such as ZigBee, Bluetooth, ad hoc Wi-Fi, etc. 
Some hosts within the organization may also support SRC protocols, which could allow the adversary to hop between networks.

\begin{definition} \textbf{Short-Range Communication.}

We define a set of short-range communication protocols $SRC = \{p_1, p_2,\ldots\}$. 
Let $src:D\cup H\rightarrow 2^{SRC}$ be a function that maps an IoT device or a host to the subset of SRC protocols that it supports. 
\end{definition}
In the remainder of this work we will use the term \emph{device} to refer to both IoT devices and hosts. 

Any two devices connected via a SRC protocol must reside within a certain distance from each other (i.e., the communication range). 
For example, let $d\in D$ be some IoT device that supports SRC protocol $p\in SRC$, and let $h\in H$ be some host that supports the same protocol.  
If $d$ is deployed in location $l$ and $h$ resides within the communication range of $l$, then $d$ may communicate with $h$ and vice~versa. 

\begin{definition}\textbf{Location Range.}
\label{def:range}

We define the $range:L\cup\{\bot\}\rightarrow 2^{D\cup H}$ 
of a particular location as the set of hosts that may communicate with an IoT device deployed there. 
\end{definition}

It is important to note that $range(l), l\in L$ is an estimation based on the radio specification of different IoT devices. 
The actual set of devices in range of IoT device deploy in location $l$ may vary depending on the power of the radio, obstacles, interference, etc.

\new{For the ease of discussion we did an abstraction to the range. We defined three different ranges: $Short Range$, $Medium Range$ and $Long Range$. 
Two devices may have the same short range communication protocol, but have different ranges. 
Each IoT device was assigned a range, depends on his $SRC$.
For each location, for each $SRC$, we defined the set of $hosts$ than are in a range from that location $l$. In case an IoT device is deployed in a location, it can connect to all hosts in that location that are with the same $SRC$ and with the same range.
Please note that the set of hosts in $Short Range$ is a subset of $Medium Range$ which is subset of $Long Range$.}


Also note that a device can be in the range of several locations and that no devices are in the range of the non-location $\bot$ (i.e. $range(\bot)=\emptyset$). 

\section{Attack Graph Definition}
\label{sec:ag_def}

The potential locations of IoT devices and $SRC$ protocols are integrated in the attack graph analysis methodology after the scanning stage and before attack graph modeling.   
For every possible deployment of IoT devices, that will be considered during the course of the optimization, we augment the connectivity map of \emph{devices} to include the hypothetical connections between any IoT device $d\in D$ deployed in location $depl(d)$ and all devices in the range of $d$ : $range(depl(d))$. 

Once the connectivity between all devices has been defined, we use the standard MulVAL framework to generate an attack graph that considers some given deployment of IoT devices. Each deployment has a different attack graph, depending on the devices deployed. If no IoT device is deployed the deployment is empty ($depl_{empty}$), and the attack graph is simply the original attack graph of the organization.

We adopt the attack graph definition introduced by Ou \etal~\cite{ou2006scalable}.

\begin{definition}\textbf{Logical Attack Graph.}
\label{def:ag}

Let $depl$ be a deployment of IoT devices in an organization. 
The logical attack graph $G_{depl}$ is a tuple:
$$G_{depl}=(N_p, N_e, N_f, E, M, g),$$
where $N_p$, $N_e$, and $N_f$ are the sets of privilege nodes, exploit nodes and fact (leaf) nodes, respectively, and $E$ is a set of directed edges
$$E\subseteq (N_p\times N_e) \cup (N_e\times (N_e\cup N_f)),$$
\end{definition}

There are two types of edges in an attack graph. 
An edge $(e,p) \in E$  from an exploit node $e\in N_e$ to a privilege node $p\in N_p$ means that the attacker can gain privilege $p$ by executing exploit $e$. 
In order to gain a privilege, an attacker needs to execute one of the exploits leading to it.

An edge $(f,e) \in E$ from a fact node or a privilege node $f\in N_f\cup N_p$ to an exploit node $e\in N_e$ means that the node $f$ is a precondition to executing the exploit $e$. 
For example, a fact node could be a vulnerability in the Bluetooth protocol that can be exploited if the attacker is in the Bluetooth range of the vulnerable device.
In order to execute an exploit, the attacker needs all of the privileges and facts that lead to the exploit.

In this work, in contrast to the definition introduced by Ou \etal~\cite{ou2006scalable}, the edge orientations follow the direction of the implied logical operation.

Next, we define the term \emph{attack plan}. For that purpose, we changed the notations from Gefen \etal~\cite{gefen2012pruning} slightly, as follows:

$pre(e) = \{ v \in N_p \cup N_f | (v,e) \in E \}$ are all of the preconditions of node $e$.

$obt(p) = \{ e \in N_e | v \in N_p \& (e,v) \in E \}$ is the set of exploits that lead to privilege node $p$ (the set of privileges the attacker obtained).

An attack plan is a sub-graph $G_{depl}'$ of some attack graph $G_{depl}$ that represents a scenario in which the attacker manages to reach the goal, namely $g \in G_{depl}'$.
Therefore, in an attack plan all of the preconditions of an exploit $e\in G_{depl}'$ are satisfied, and each privilege $p\in G_{depl}'$ is obtained by an exploit.

\begin{definition}\textbf{Attack Plan.}

Let $AP(G_{depl})$ be all of the attack plans of graph $G_{depl}$.
Each attack plan $G_{depl}'\in AP(G_{depl})$ needs to satisfy these three conditions:
\begin{itemize}
    \item $g \in G_{depl}'$
    \item $\forall a \in N_e : pre(a) \subseteq G_{depl}' | N_e \in G_{depl}' $
    \item $\forall p \in N_p : \exists a \in obt(p) \subseteq G_{depl}' | N_p \in G_{depl}' $
\end{itemize}

We consider the length of an attack plan as the number of nodes it contains.
$OptLen(G_{depl})$ is the length of the shortest attack plan in graph $G$, a $OptCnt(G_{depl})$ indicates how many of the shortest attack plans there are in graph $G$.
\new{We also consider the average number of exploits in the shortest attack plans in graph $G$, as $OptExp(G_{depl})$, and the average number of privileges in the shortest attack plans, as $OptPrv(G_{depl})$.
}

\end{definition}


\section{Risk Score}
\label{sec:risk_score}
%

The network security can be estimated by the Risk Score, where the higher the risk score the lower the security of the network.
In an environment in which IoT devices are deployed, there are a few aspects to consider when choosing a method for computing the risk score. 

First, the method needs to convey that the deployment of IoT devices may generate new attack plans.
Consequently, the cost of an attack may drop and the likelihood of an attack may increase due to the additional vulnerabilities and opportunities for lateral movement that an attacker can exploit.
Last, the method needs to indicate the changes in different deployments and be sensitive enough to detect the changes caused by the deployment of even a single additional IoT device.

We did not find any risk score in the literature that meet these requirements. Most risk scores are specific for a problem, they are trying to solve some problem that was not taken into consideration in other works (zero-days, for example).

As we will see, our work is designed for all risk scores, as long as they are monotonic. Meaning that deploying a device could not reduce the risk.

We consider a deployment of IoT devices that combine several aspects of the network. We choose to calculate all the shortest attack plans, taking into consideration their length, quantity, average number of exploits and average number of privileges. 
Our main goal is to compare between different deployments, while trying to determine between two networks which one is more secure. To do that, we calculated all the above parameters for the network with no IoT devices deployed, and the risk score of each deployment is derived from that.

Our purpose was to find a method that will allows us to compare between different deployments. Due to the fact that we have a baseline, which is a deployment with no IoT device deployed, our risk score is a comparison to that.

\remove{
We consider a deployment of IoT devices that reduces the number of options the attacker has for an attack. 
Therefore, in our work, we choose to calculate the shortest attack plans, taking their length and quantity into consideration. }
Gonda \etal~\cite{gonda2017ranking} describes the computation of the shortest attack plans in detail. 
As noted by the authors, enumerating all of the attack plans is NP-hard, which means that the running time can be exponential, however, we performed this computation on several networks, and the running time was short, as can also be seen in Section~\ref{sec:results}.

\begin{definition}\textbf{Risk Score.}

\new{
$R(depl)$ is a number that represents the risk score of deployment $depl$, and $R(depl_{empty})$ is the risk score of the deployment with no IoT devices deployed.}

\new{
$$R(depl) = 
OptLen(G_{depl}) / OptLen(G_{depl_{empty}}) - 1 +$$ 
$$OptCnt(G_{depl}) / OptCnt(G_{depl_{empty}}) - 1 +$$
$$OptExp(G_{depl}) / OptExp(G_{depl_{empty}}) - 1 + $$
$$OptPrv(G_{depl}) / OptPrv(G_{depl_{empty}}) - 1 $$ 
}

\remove{$R(depl)$ is a tuple that represents the risk score of deployment $depl$.
The first element is the length of the shortest attack plan in graph $G_{depl}$, and the second element indicates how many of the shortest attack plans there are.
$$R(depl) = (OptLen(G_{depl})), OptCnt(G_{depl}))$$
}

\end{definition}

\new{In short, the risk score is the total of the relative increasing of each parameter. Note that $R(depl_{empty}) = 0$. 
The minimum value of $R(depl)$ is zero, the same as an empty deployment. A value of one means that on average, the value of each parameter has doubled.}

\chapter{Deployment Optimization Problem}
\label{sec:deployment_opt}
In this section, we introduce the terms and notation used to define the two IoT deployment optimization problems: (1) Full Deployment with Minimal Risk (FDMR), and (2) Maximal Utility without Risk Deterioration (MURD).

\subsection{FDMR Problem} 
Given an attack graph of an organization $G$, a set of IoT devices $D$ of types $T$, and the location constraints $C$, 
find the deployment ($depl_{full}$) of IoT devices such that all of the IoT devices are deployed subject to location constraints, and the risk score $R(depl_{full})$ is minimized.  

\begin{definition}\textbf{Full Deployment with Minimal Risk (FDMR) Problem.}

Given the four-tuple $<G,D,T,C>$, find $depl_{full}$ such that $R(depl_{full})$ is minimized 
$$\argmin_{depl_{full}}\{R(depl_{full})\}$$
\end{definition}

\subsection{MURD Problem}
Given an attack graph of an organization $G$, a set of IoT devices $D$ of types $T$, and the location constraints $C$, find the deployment that consists of the highest number of IoT devices without increasing the risk score $R$. 

\begin{definition}\textbf{Maximal Utility without Risk Deterioration (MURD) Problem.}

Given the four-tuple $<G,D,T,C>$, find $depl$ such that $|R(depl)|$ is maximized and $R(depl)$ = $R(depl_{empty})$ 
$$\argmax_{depl}\{|R(depl)| : R(depl) = R(depl_{empty})\}$$
\end{definition}

\section{Search Space}

Next, we define the search space for both FDMR and MURD. 
In each case, the state of the search space is organized as a binary tree where at each state a decision is made either to deploy (left child) or not to deploy (right child) a particular IoT device in a particular location. 
The root state is an empty deployment ($R(depl_{empty})$) where no decisions have been made yet. 
Every path from the root node of the search space corresponds to a set of decisions. 
This means that a path from the root to any state defines where some of the IoT devices are deployed and where some other IoT devices cannot be deployed.
The set of left children along a path is a partial deployment of IoT devices. 
In this way, we consider all possible deployments, subject to location constraints.

For every node of the search space we derive the respective attack graph $G_{depl}$ and compute the risk score $R(depl)$. 
The goal nodes depend on the specific problem. 
In the FDMR problem the goal nodes include all states with a deployment that meets all of the constraints, $depl_{full}$, and the objective is to identify the goal state with the lowest risk score.
In the MURD problem the goal states include all states with a deployment that has the same risk score as the initial state.


\new{
\subsection{Search Space Size}
\label{sec:search_space_size}
Using the definitions above, we calculate the size of the search space, meaning the number of possible full deployments.
We find the total number of the different deployments depending on the existing constraints by calculating the number of possible deployments for each location type, and multiplying all the results with each other.
To calculate the number of all possible deployments in $t$ we use permutation~\footnote{Permutation $_nP_k$ means that for $n$ items, we want to find the number of ways $k$ items can be ordered.} and combination~\footnote{Combination $\binom nk$ is a selection of $k$ items from a collection of size $n$, such that the order of selection does not matter.}. 
Permutation is define as: 
$_nP_k=\frac{n!}{(n-k)!}$,
and combination is define as:
$\binom nk=\frac{n!}{k!(n-k)!}$.
}

\new{In the following explanation, we will use the example in Definition~\ref{def:locations_const} to
calculate the number of possible deployments of type TV, 
}

\new{
$C(TV) = (\{l_{TV_1},l_{TV_2},l_{TV_3}\},2, \{d_{tv1},d_{tv2},d_{tv3},d_{tv4}\}$.
}

\new{
First, we calculate the permutation of number of devices there are in $t$ type ($|D(TV)|=4$) out of locations to deploy ($n(TV)=2$).
We use permutation because in this case the order matter, since each device is unique and the location of each device means different deployment. 
If we choose two devices and deploy them in two locations, and if we choose the same two devices but switch the locations, the deployments would be different, $_{|D(TV)|}P_{n(TV)} = {_4}P_2 = 12$.
Next, we calculate the combination of the number of locations that needed to be deployed in type $t$ ($n(TV)=2$) out of locations to deploy ($|L(TV)|=3$).
In this case we use combination because the order does not matter. There are three locations in total and we need to deploy only two of them.
$\binom{|L(TV)|}{n(TV)} = \binom{3}{2} = 3$.
}

\new{
Therefore, the number of possible different deployments in type $TV$ is $36$, $3\cdot 12=36$.
The formula to compute the size of possible deployments, which is also define as the size of the search space, is as follows: 
}

\new{
$$\Pi_{t\in T}\left(\binom{|L(t)|}{n(t)}\cdot _{|D(t)|}P_{n(t)} \right)$$
}

\new{
We multiple the number of all possible deployments of each type.
}

\section{Search Algorithm}
\label{sec:heuristic}
For our heuristic search, we used the DFBnB algorithm (as described in Section~\ref{sec:back-search}). The heuristic function will be described later in this section.
We choose DFBnB algorithm because it's pruning ability can significantly reduce that running time. In addition, in the way we built the search space, our branching factor is two (each node has exactly two children), and the algorithm is suitable for problems with low branching factor. 
\new{A pseudo code of the algorithm can be found at Algorithm~\ref{alg:dfbnb} and is further explained in Section~\ref{sec:pseudo_code}. }

As we mentioned above, each state in our search tree has two children (left and right). 
In one, we added an IoT device to the deployment in a certain location, and in the other, we did not allow the IoT device to be deployed in that location. 
In practice, each state has various options regarding which IoT devices to deploy. 
We \new{used node ordering to } chose one device ($d$) and one location ($l$) where $d$ can still be deployed and generate two children: deploy $d$ at $l$ and do not deploy $d$ at $l$.
For the left child corresponding to the deploy decision, we generate a new attack graph and recalculate the risk score 
.
We do not calculate the risk score for the right (do not deploy) child, as this child's risk score did not change, since the risk score depends only on the deployed devices.

\new{
\subsection{Pseudo-Code}
}
\label{sec:pseudo_code}
\new{
At first, we initialized our variables, including adding the empty deployment, with no IoT devices deployed, to a stack ($root$). For each state we popped from the stack, we generated two children, as described in Section~\ref{sec:heuristic}. $s_l$ is corresponding to the left generated child, while $s_r$ is corresponding to the right. We calculated the $f$-value for each child (note that if there is no heuristic function, $f(x)=g(x)$). The security risk calculation is based only on the left child. When we reach a goal node (full deployment), we save it as our best solution and update $alpha$. We add to the stack a child only if it's $f$-value is smaller than $alpha$.}

\new{
Please note that the showed pseudo-code is suitable for the FDMR problem. The pseudo code for the MURD problem is the same, except for the signs in lines $13,15$, which are the opposite ($\ge$ instead of $\le$).
}

\begin{algorithm}
\caption{DFBnB algorithm}\label{alg:dfbnb}
\begin{algorithmic}[1]
\Procedure{DFBnB}{$a,b$}
\State $alpha\gets$ $\infty$
\State $bestSolution\gets$ null
\State $stack\gets$ $\{root\}$ 

\While{$stack\not=0$}
\State $state\gets stack.pop$
\State $s_l, s_r\gets getTwoSons(state)$
\State $fs_{l}, fs_{r} \gets calcF(s_{l}, s_{r})$
\State $r \gets calcSecurityRisk(s_{l})$

\If{$isGoal(s_l)$}
\State $alpha\gets f{s_{l}}$
\State $bestSolution \gets r$
\EndIf

\If{$fs_{1}$ $\le$ $alpha$}
\State $stack\gets$ $\{s_{l}\}$ 
\EndIf

\If{$fs_{r}$ $\le$ $alpha$}
\State $stack\gets$ $\{s_{r}\}$ 
\EndIf

\EndWhile
\State \textbf{return} $bestSolution$
\EndProcedure
\end{algorithmic}
\end{algorithm}

\subsection{Heuristic Functions}
In order to calculate the heuristic functions, we created a table of risk scores $Table(depl_n)$ which contains the risk scores for each IoT device in each possible location. 
In other words, we simulate the deployment of a single IoT device each time. 
For each deployment, we update the table, removing the IoT device that was deployed or not allowed to be deployed. 

\begin{definition}\textbf{FDMR Heuristic Function.}

For the FDMR problem, the heuristic function underestimates the lowest possible change in risk in every subtree.
Then, whenever the risk score of the best full deployment found so far is lower than the risk score of any full deployment that can be found within a subtree, that subtree is pruned.

For FDMR, let $h_{FDMR}(depl_n)$ be the heuristic of $depl_n$.
$h_{FDMR}(depl_n)$ is the minimal $R(depl_d)$ 
and $R(depl_d) \in Table(depl_n)$.
$$ h_{FDMR}(depl_n) = \argmin_{d\in D}\{R(depl_d) \in Table(depl_n) \}$$
Intuitively, $h_{FDMR}$ underestimates the risk score because (1) individually each deployed device increases the risk according to $Table(depl_n)$, but (2) together multiple deployed devices may result in attack plans that were not accounted for yet.   
\end{definition}

\begin{definition}\textbf{MURD Heuristic Function.}

For the MURD problem, the heuristic function overestimates the highest possible change in the number of IoT devices that can be deployed without increasing the risk. 
Then, whenever the number of devices deployed according to the incumbent solution found so far is larger than the number of devices that can possibly be deployed by continuing to search a subtree, that subtree is pruned.

We want to deploy the highest number of IoT devices possible, hence the heuristic function counts the number of IoT devices in $Table(depl_n)$ with the same risk score as the root state.
Let $h(depl_n)$ be the heuristic of $depl_n$. $h_{MURD}(depl_n)$ is the number of devices with a risk score equal to initial state $R(depl_{empty})$, such that 

$|R(depl_d)=R(depl_{empty})|$ and $R(depl_d) \in Table(depl_n)$.
$$ h_{MURD}(depl_n) = |R(depl_d) \in Table(depl_n) : R(depl_d)=R(depl_{empty})|$$
Intuitively, $h_{MURD}$ overestimates the number of devices that can be deployed because (1) any IoT device that increases the risk according to $Table(depl_n)$ cannot be deployed, and (2) even if individually a set of deployed devices does not increase the risk score, together they may result in an attack plan that was not available before.   
\end{definition}

\new{
\subsection{Node Ordering}
In order to find an optimal goal node quickly, the newly generated child nodes should be searched in an increasing order of their costs. This method is called \textit{node-ordering}, and can significantly speed up the search~\cite{reinefeld1994enhanced, zhang1995performance}.
In our case, we used our heuristic function also as node ordering. In other words, we generates the child with the lowest heuristic. When we didn't use heuristic function, we randomly chose a child to generate.
}



\chapter{Experiments}
\label{sec:evaluation}
We conducted experiment for each one of the problems we wish to solve: finding the full deployment with minimal risk (FDMR), and finding the maximal utility without risk deterioration (MURD).
For both problems, we used the suggested DFBnB algorithm with the heuristics described in Section~\ref{sec:heuristic}.

\section{Data Preparation}
\label{sec:data}
To evaluate our proposed method, we conducted a set of experiments using an attack graph that was derived from a real organization network.


\subsection{Organization Network}
\begin{figure}[ht]
\centering
\includegraphics[width=0.75\textwidth]
{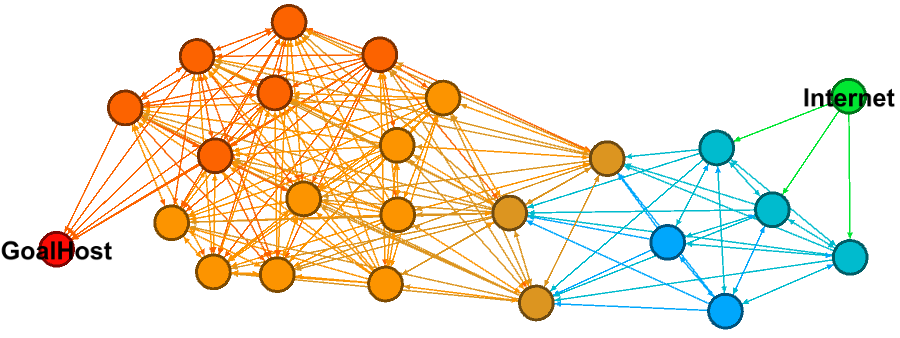}
\caption{Connectivity graph of the hosts in the organizational network, derived from the VLAN topology. The different colors represent the different VLANs. The blue nodes are DMZ VLAN, and the orange nodes are the internal organization network. Each node represents a host, and an edge indicates a connection between two hosts.}
\label{fig:org_graph}
\end{figure}

The network of the organization is a real network consisting of $24$ hosts which was used by Gonda \etal~\cite{gonda2017ranking}.
The network of the organization was scanned using Nessus Scanner, and then MulVAL was used to generate the attack graph based on the scanning results.
Figure~\ref{fig:org_graph} depicts the connectivity of the hosts in the network, derived from the VLAN topology. Each node represents a host, and an edge indicates a connection between two hosts.

An organization can have more than one host that it wishes to protect, and this is translated to multiple targets for the attacker.
To simplify things, all target hosts are connected to an abstract $goalHost$, and the goal of the attack graph is to execute code in this host.
Executing code on the $goalHost$ proves that the attacker managed to control one of the targeted hosts that led to the goal.
As part of the experimental setup we assume that the organization is free from inside adversaries and that the potential attacker is located on the $Internet$.
The attack graph has a host that represents the Internet.
Detailed information on the scanning process is provided in~\cite{gonda2017ranking}.

\subsection{Simulating IoT Devices}
The network of the organization used in the experiment does not include any IoT devices.
Therefore, we opt to simulate the IoT devices, their communication protocols, and the constraints required for their deployment.
We simulated three IoT types (detector, refrigerator, camera), nine different IoT devices (four detectors, two cameras, and three refrigerators), and eight locations for the deployment of IoT devices.

In the simulation, the organization would like to deploy three detectors for which there are four possible locations, one camera for which there are two possible locations, and two refrigerators for which there are two possible locations. 
Therefore, a total of six IoT devices needed to be deployed.

Formally, as defined in Definition~\ref{def:locations_const}, the location constraints in our simulation are defined as follows:

$C(detector)$ = $(\{l_{det_1},l_{det_2},l_{det_3}, l_{det_4}\}$,$3$,$\{d_{det1},d_{det2},d_{det3}$,$d_{det4}\})$

$C(camera) = (\{l_{cam_1},l_{cam_2}\},1,\{d_{cam1},d_{cam2}\})$

$C(refrigerator) = (\{l_{ref_1},l_{ref_2}\},2,\{d_{ref1},d_{ref2},d_{ref3}\})$

\remove{
Using permutation\footnote{Permutation $P^n_k$ mean that for $n$ items, we want to find the number of ways $k$ items can be ordered.}
and combination,\footnote{Combination $C^n_k$ is a selection of $k$ items from a collection of size $n$, such that the order of selection does not matter.} we can calculate the total number of options in the search space.
}

\new{We can calculate the size of the search space (See Section~\ref{sec:search_space_size} for explanation).} 


$ \binom 43 {_4}P_3 \cdot \binom 21 {_2}P_1 \cdot \binom 22 {_3}P_2 = 2304$

Meaning, there are $2304$ possible deployments.


\subsection{Simulating Short-Range Communication} 
We simulated two short-range communication protocols (ZigBee and Bluetooth) and randomly divided them between all IoT devices and hosts so that 75\% of the hosts have Bluetooth and 20\% number of them have Zigbee, and 40\% of the IoT devices have Bluetooth and 80\% of them have Zigbee. 
\new{The chance for a Zigbee is not dependent on the chance of Bluetooth, and vice versa. }

\new{For each device for each of his short range communications protocols, we randomly choose the range of the protocol. Usually, Bluetooth devices have longer range than ZigBee devices. Due to that, devices with Zigbee protocols could get only $Short Range$ (60\% chance) or $Medium Range$ (40\% chance), and devices with Bluetooth protocols could get $Medium Range$ (60\% chance) or $Long Range$ (40\% chance).}


\subsection{Simulating Vulnerabilities} In order to create potential attack plans that include IoT devices, we simulated existing vulnerabilities that can be exploited as follows. For each IoT device and for each host, in addition to its known vulnerabilities (from the scanning performed), we created a vulnerability based on the protocol used.

\subsection{Simulating Physical Location of Hosts} 
The actual physical location of the real hosts was unavailable. 
The location of the hosts is important in order to simulate the proximity of the IoT devices to the host, and consequently create potential attack plans involving the IoT devices. 
Therefore, we randomly divided the hosts among the eight simulated location ranges. 
\new{For each location, for each of the short range communications protocols, we randomly chosen hosts that are in proximity to the location, depends on the range. }
Note that a host can be in proximity to more than one IoT device.

\section{Experimental Setup}
The experiments were conducted on Hyper-V VM, with four virtual CPUs (two cores) and 8GB RAM. The setup of the experiments is as follow:

\subsection{Number of Executions} In order to strengthen the validity of our results, we executed the experiment forty times, using a different host location each time. 
In other words, we simulated the physical location of hosts forty times. 
The results in the next section are the average results of all executions.


\subsection{Evaluation Measures} 
We computed two measures: the first is the execution time, and the second is the risk score of a suggested IoT deployment (for the FDMR use case) or the number of deployable IoT devices (for the MURD use case). 
The evaluation measures were averaged over the all of the executions. 
The execution time is important, since this can be a weak point, as one of the difficulties in attack graphs and solutions that are based on attack graphs is execution time.
    
\subsection{Random Deployment} 
For comparison, we also ran both problems randomly as a baseline. 
This scenario represents an organization that randomly deploys IoT devices, without considering the security aspect. 
That is to say, for the FDMR problem we randomly deployed all IoT devices five times and took the average risk score of all the deployments.
In the MURD problem, each time we added a device randomly and computed the risk score. 
We started with no IoT devices deployed and continued until full deployment. We ran five times each number of devices. 
This random baseline was executed the same number of times as our algorithm (forty times).  

\chapter{Results}
\label{sec:results}

\new{
Table~\ref{tab:results_algo} and~\ref{tab:results_heuristic} present the results for both problems, FDMR and MURD.
Table~\ref{tab:results_algo} includes the results of the DFBnB algorithm and random deployment. Table~\ref{tab:results_heuristic} includes a comparison between using the DFBnB algorithm with and without heuristic function.
Table~\ref{tab:risk_stats} presents the average value of each of the risk score components. The quantity of the shortest paths is the parameter that affect the risk score the most. The length of the shortest paths and the number of privileges they contain did not change between different risk scores.
}

\begin{table}[ht]
\centering
\begin{tabular}{@{}|l|c|c|c|c|@{}}
\toprule
      & \multicolumn{1}{l|}{Exploits} & \multicolumn{1}{l|}{Privileges} & \multicolumn{1}{l|}{Shortest Paths Count} & \multicolumn{1}{l|}{Shortest Paths Length} \\ \midrule
DFBnB  & 2.8                           & 8                          & 1311                       & 29                          \\ \midrule
Random & 2.5                           & 8                          & 1854                       & 29                          \\ \bottomrule
\end{tabular}
\caption{DFBnB and Random Risk Scores Stats Distribution}
\label{tab:risk_stats}
\end{table}

\new{
\section{DFBnB algorithm Vs Random Deployment}
\textbf{Full Deployment with Minimal Risk (FDMR).}
In the FDMR problem, the average risk score of all runs is $0.22$, a 22\% increase compare to the initial state. 
The algorithm took an average of $8.48$ minutes to run, which is a reasonable amount of time and provides an indication of its feasibility on a larger scale.
}

\begin{table}[ht]
\centering
\resizebox{\textwidth}{!}{%
\begin{tabular}{@{}|c|c|c|c|c|c|@{}}
\toprule
\multirow{2}{*}{Problem} & \multicolumn{3}{c|}{DFBnB}                             & \multicolumn{2}{c|}{Random}               \\ \cmidrule(l){2-6} 
                         & Risk Score (std) & Deviced Deployed (std) & Time (min) & Risk Score (std) & Deviced Deployed (std) \\ \midrule
FDMR                     & 0.22 (0.22)      & 6 (0)                  & 8.48       & 0.64 (0.37)      & 6                      \\ \midrule
MURD                     & 0                & 4.4 (1.28)             & 3.23       & 0                & later                  \\ \bottomrule
\end{tabular}%
}
\caption{Comparison between DFBnB algorithm and Random deployment (average over 40 executions).}
\label{tab:results_algo}
\end{table}

\new{
\noindent\textbf{Maximal Utility without Risk Deterioration (MURD).}
In the MURD problem, the average number of IoT devices that can be deployed without affecting the security risk is $4.40$. This number means that, on average, four to five devices can be deployed without any change in the risk score.
It took the algorithm an average of $3.23$ minutes to compute, which is also a reasonable time.
}

\new{
\noindent\textbf{Random Deployment.}
In FDMR, the average risk score was $0.64$, an increase of 64\% from the initial state. We can see that randomly deploying IoT devices leads to less safe network, compared to the increase of only 22\% when using our algorithm.
}

\new{
In the MURD problem, the average risk score of deploying four IoT devices is $0.41$. Meaning when deploying four devices randomly, the risk score increases by $41\%$. We chose four devices because with our algorithm we managed to deploy an average of $4.40$ devices without influencing the security of the network.
The average risk score of other numbers of devices can be seen in Figure ~\ref{fig:murd_stats}~(in grey), where we present the average risk score of deployments with each number of devices, ranging from zero (empty deployment) to six (full deployment).}

\new{
\section{With Vs Without Heuristic}
\textbf{Full Deployment with Minimal Risk (FDMR).}
With the heuristic function, the average risk score for FDMR problem is about eight and a half minutes. Without heuristic, the time increases to almost $150$ minutes, more than $17$ times compare to using the heuristic function. We can see that in both cases the result is the same.
}

\new{
\begin{table}[h]
\resizebox{\textwidth}{!}{%
\begin{tabular}{@{}|c|c|c|c|c|@{}}
\toprule
\multirow{2}{*}{Problem} & \multicolumn{2}{c|}{Time (min)}    & \multicolumn{2}{c|}{Risk Score}    \\ \cmidrule(l){2-5} 
                         & With Heuristic & Without Heuristic & With Heuristic & Without Heuristic \\ \midrule
FDMR                     & 8.48           & 148.83            & 0.21           & 0.21              \\ \midrule
MURD                     & 3.23           & 0.3               & 4.40           & 4.40              \\ \bottomrule
\end{tabular}%
}
\caption{Results: With and Without Heuristic (average over 40 executions)}
\label{tab:results_heuristic}
\end{table}
}

\new{
\noindent\textbf{Maximal Utility without Risk Deterioration (MURD).}
As can be seen, not using the heuristic function is ten times faster than using it. The reason for this is due to the extra time using a heuristic function is adding. On average, it take about $45$ seconds to compute the heuristic table. Also, using a heuristic is another operation that may take some time, as we need to look in the table and update it. In this case, using the heuristic did not helped in improving the running time, and the algorithm did well on his own without it.
}

\section{Additional Results}

\subsection{Trade-off}
\begin{figure}[ht]
\centering
\includegraphics[width=1\textwidth]
{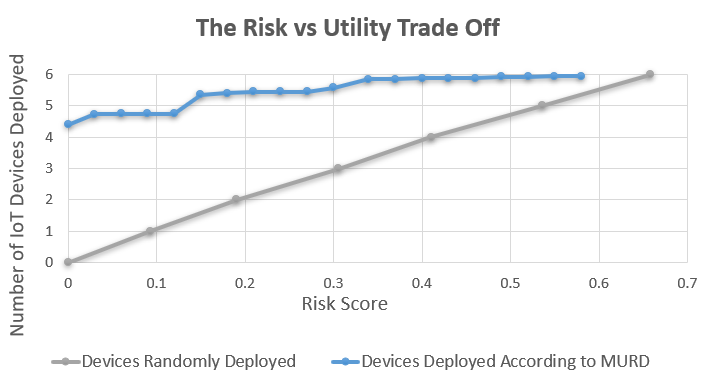}
\caption{The blue graph indicates the average number of devices deployed under security risk bound. The grey graph indicates the average risk score of deployments with each number of devices. 
} 
\label{fig:murd_stats}
\end{figure}

We investigated the trade-off between the allowed risk of the IoT deployment and the maximal number of IoT devices that can be deployed. 
Figure~\ref{fig:murd_stats} further emphasizes the difference between random and optimal deployment of IoT devices.  
On one hand, 4-5 randomly deployed IoT devices increase the security risk by 40\%. 
On the other hand the same number of IoT devices can be deployed with insignificant risk deterioration.      
We can also see from Figure~\ref{fig:murd_stats} that the difference between optimal and random deployment strategies diminishes as we try to deploy six IoT devices. 

\subsection{Cumulative Distribution}
\begin{figure}[ht]
\centering
\includegraphics[width=1\textwidth]
{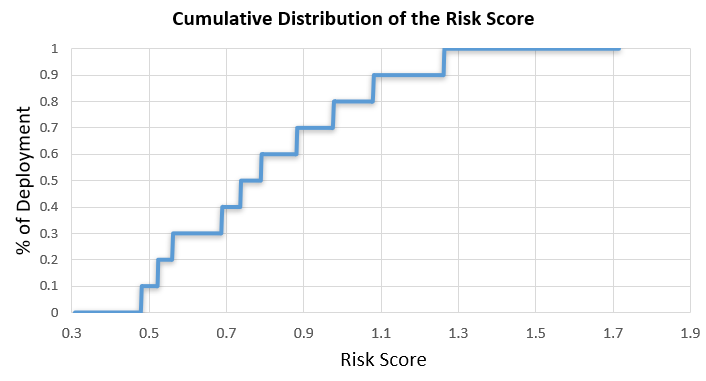}
\caption{The $x$-axis is the cumulative risk score, and the $y$-axis is the percentage of deployments for which the risk score is less than $x$.
} 
\label{fig:dist}
\end{figure}

Figure~\ref{fig:dist} illustrates the challenge in finding the safest deployment of IoT devices.
The graph presents the cumulative distribution of the risk scores of all deployments in one execution. 
The $x$-axis is the cumulative risk score, and the $y$-axis is the percentage of deployments for which the risk score is less than $x$. 
As can be seen, 50\% of the deployments have a risk score lower than \new{$0.78$}. 
Moreover, only $12$ deployments (\new{0.04\% } of all deployments) are optimal, with a risk score of \new{$0.307$}, i.e., the chances of a random selection to choose an optimal deployment in that execution was \new{$0.00004$}.

\subsection{Robustness}
The risk score of an optimal deployment may change when new vulnerabilities are discovered, leading to potentially inferior deployment. 
To conclude the experimental evaluation we tested the robustness of the optimal deployment
of an arbitrary execution from the FDMR problem, with risk score of \new{$0.3164$}. 
We perturbed vulnerabilities of 10\% and 20\% of the devices in the network by discarding all current vulnerabilities of the chosen devices and randomly assigning new vulnerabilities as described in Section~\ref{sec:data}. 
This process was repeated ten times. 
The average risk score of the optimal deployment after changing 10\% of the vulnerabilities \new{was $0.308$ with standard deviation of $0.007$. For changing 20\% of the vulnerabilities, the average risk score was $0.266$ with standard deviation of $0.07$. } 
Overall the changes in the risk of the optimal deployment due to perturbation of the vulnerabilities were not statistically significant. 
\chapter{Conclusion}
\label{sec:conculsion}
We present a novel method for suggesting the optimal deployment (in terms of the security risk) of a set of IoT devices within an organization. 
In order to accomplish this, we augmented the conventional attack graph to include short-range communication protocols inherent to IoT devices.
To the best of our knowledge, this is the first work that takes the physical location of devices and different communication protocols into account.

We demonstrated the importance of planning a deployment of IoT devices by solving two scenarios, approaching them as an optimization problem. 
We proposed a novel method for evaluating the risk of IoT device deployment using an augmented attack graph, and used the proposed method to address these two scenarios. 
Our results revealed the potential risk in deploying IoT devices in organizations and showed that randomly deploying devices can greatly affect the security of the organization's network. 
We solved the two scenarios on a real organization with a small to medium sized network, with a running time of less than ten minutes.

\chapter{Future Work}
\label{sec:future_work}
Our algorithm, and in particular, our heuristic approach, assumes that the potential risk of two deployed devices is greater than or equal to the sum of their individual risk scores. Any method of risk calculation that satisfies this assumption can be used in the algorithm. 
The method of risk score calculation used in this work has some limitations. It does not take the cost of different exploits into account, which can be a major consideration for an attacker. As a result, the method does not capture the heterogeneity and homogeneity of vulnerabilities along an attack path. 

Future work may extend the current research in the following directions.
First, it is desirable to increase sizes of the attack graph that can be optimized by providing more accurate heuristic functions. 
In addition, the optimization methods proposed in this work should be tested with variety of risk scores that encompass the true cost of the attack, the probability of the attack success, or both. 
Finally, cyber-physical capabilities of IoT devices as well as their unique functionalities should be incorporated into an extended model.